\documentclass[%
 reprint,
 amsmath,amssymb,
 aps,
]{revtex4-2}

\usepackage{graphicx}
\usepackage{dcolumn}
\usepackage{bm}

\usepackage{hyperref}
\usepackage{braket}
\usepackage{dsfont}
\usepackage{float}
\usepackage{breqn}
\usepackage{esint}
\usepackage{xcolor}
\usepackage{notoccite}
\usepackage{physics}
\newcommand{\comma}{\char`,{} }
\begin{document}

\preprint{APS/123-QED}
\title{Real space circuit complexity as a probe of phase diagrams} 

\author{Nishan C. Jayarama}
\email{nishancj@iitk.ac.in}
 \affiliation{Department of Physics\comma Indian Institute of Technology\comma Kanpur\comma 208016\comma Uttar Pradesh\comma India}
\author{Viktor Svensson}%
 \email{viktor.svensson@ftf.lth.se}
\affiliation{%
Division of Solid State Physics and NanoLund\comma Lund University\comma S-22100 Lund\comma Sweden }%

\date{\today}

\begin{abstract}
Circuit complexity has been used as a tool to study various properties in condensed matter systems, in particular as a way to probe the phase diagram. However, compared with measures based on entanglement, complexity has been found lacking. We show that when imposing penalty factors punishing non-locality, it becomes a much stronger probe of the phase diagram, able to probe more subtle features. We do this by deriving analytical solutions for the complexity in the XY chain with transverse field. 
\end{abstract}

\maketitle


\section{Introduction and Outline}
The borrowing of tools from quantum information science to study models in different areas of physics has been a fruitful undertaking, particularly in high-energy physics and condensed matter systems. One of these tools is the computational complexity - the minimum number of elementary operations to complete a given task. 
While initially applied to studies of algorithms, a geometric formulation of complexity \citep{nielsenGeometricApproachQuantum2005,nielsenQuantumComputationGeometry2006, dowlingGeometryQuantumComputation2006, nielsenOptimalControlGeometry2006} has in recent years found use in several other fields.
In the AdS/CFT correspondence \cite{maldacenaLargeLimitSuperconformal1999}, entanglement entropy is dual to the area of surfaces \cite{ryuHolographicDerivationEntanglement2006}. The proposal that complexity plays the role of something like a volume \cite{susskindEntanglementNotEnough2014, susskindComputationalComplexityBlack2014, susskindAddendumComputationalComplexity2014, stanfordComplexityShockWave2014, brownComplexityActionBlack2016, brownComplexityEqualsAction2016} has generated a lot of activity in defining and calculating new observables on both sides of the correspondence. Many papers have studied this on the quantum field theory side, developing the machinery and applying it to a variety of systems
\citep{chapmanComplexityQuantumField2018a,jeffersonCircuitComplexityQuantum2017a, yangComplexityQuantumField2018, guoCircuitComplexityCoherent2018a, alvesEvolutionComplexityFollowing2018, chapmanComplexityEntanglementThermofield2019b, hacklCircuitComplexityFree2018a, khanCircuitComplexityFermionic2018a, jiangCircuitComplexityFree2020, yangPrinciplesSymmetriesComplexity2019,  yangMoreComplexityOperators2019, yangComplexityOperatorsGenerated2019, floryGeometryComplexityConformal2020a}.

Independent of its role in holographic duality, a natural question is to what extent complexity can be used as a tool to probe the properties of different systems. For example, can complexity diagnose different phases or characterize quantum chaos? A number of works have shown this to be true
\citep{liuCircuitComplexityTopological2020a, xiongNonanalyticityCircuitComplexity2020, aliPostquenchEvolutionComplexity2020a, jaiswalComplexityInformationGeometry2021b, jaiswalComplexityInformationGeometry2022, palEvolutionCircuitComplexity2022, palComplexityLipkinMeshkovGlickModel2022, aliChaosComplexityQuantum2020, camargoComplexityNovelProbe2019, bhattacharyyaCircuitComplexityInteracting2018, balasubramanianQuantumComplexityTime2020}.

\begin{figure}[b]
    \centering
    \includegraphics[width=0.35
    \textwidth]{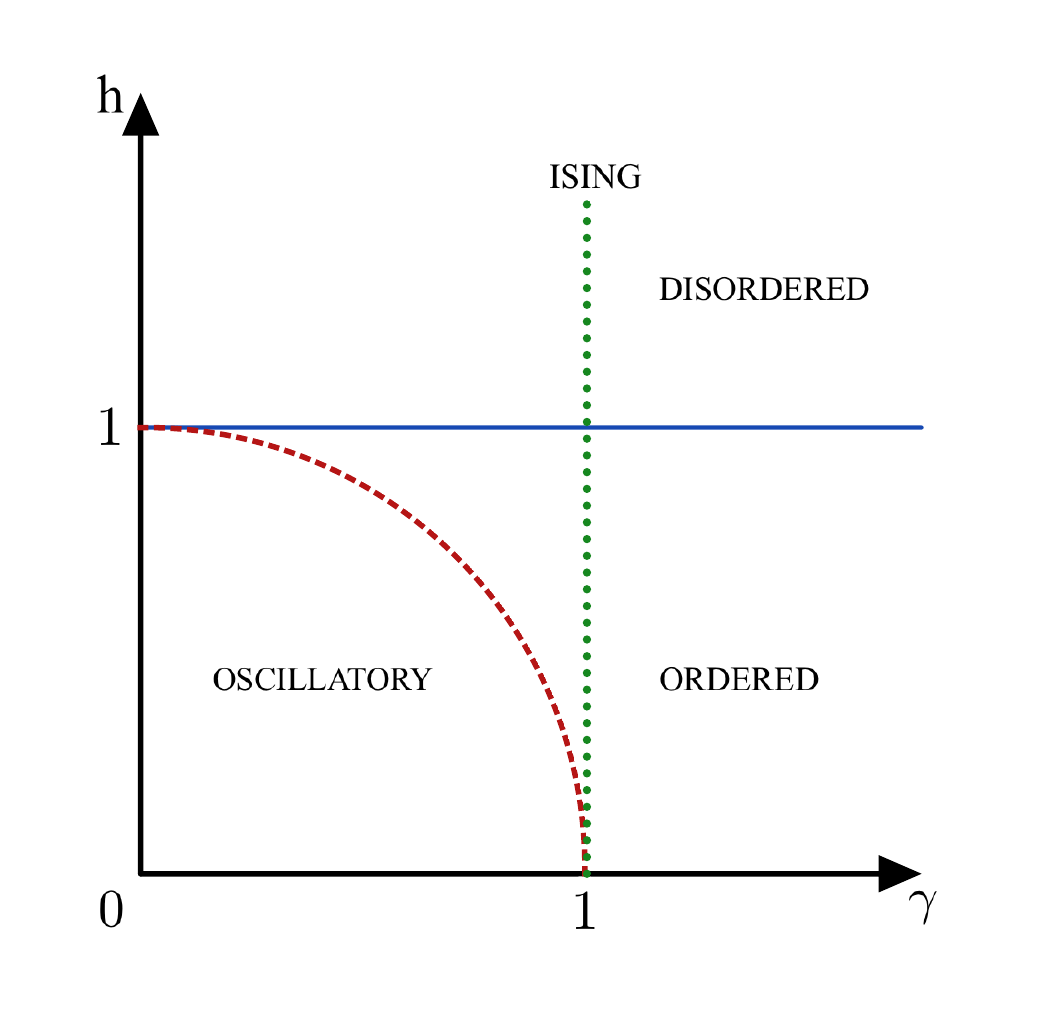}
    \caption{Phase diagram of the XY model. Adapted from \cite{franchiniIntroductionIntegrableTechniques2017a}. The blue solid line is a phase transition, which can be probed by the complexity \cite{liuCircuitComplexityTopological2020a, xiongNonanalyticityCircuitComplexity2020}. In this work, we show how a real space interpretation of the complexity enables us to detect also the red dashed curve.}
    \label{fig:phase_diagram}
\end{figure}

Much of this work is made in the context of free fermion models, where the optimal circuit implements a Bogoliubov rotation in momentum space. In the transverse field XY chain, the analytic properties of the complexity of these momentum space circuits show signatures of phase transition \cite{liuCircuitComplexityTopological2020a, xiongNonanalyticityCircuitComplexity2020}. There is another feature of the phase diagram, the red dashed curve in Fig.\,\ref{fig:phase_diagram}, called the factorizing curve. States along this curve factorize into a product state of the local spins and divide regions where correlation functions exhibit oscillatory behavior or not. It can be detected in measures based on entanglement \cite{weiGlobalEntanglementQuantum2005, amicoDivergenceEntanglementRange2006, giampaoloUniversalAspectsBehavior2013, chitovDisentanglementDisorderLines2022}, but it was not seen in studies of circuit complexity.

These studies use a circuit based on Bogoliubov rotations in momentum space, which is inherently non-local. A fully local treatment would start with some set of local gates, such as nearest neighbor hoppings and pairings, and construct an optimal circuit out of these. This notion can be used to classify topological phases \cite{chenLocalUnitaryTransformation2010} but deriving explicit circuits and the complexity is a hard problem.

In this work, we pursue another strategy of including the effects of locality on circuit complexity. We reinterpret the Bogoliubov circuit as a real space circuit, thereby changing the cost function and enabling the use of penalty factors as a way of punishing non-locality. The simpler problem that we attack can be solved analytically, but still, allows us to uncover some interesting features that we expect to show up in a fully local treatment. With these penalty factors, complexity is sensitive to more subtle features of the phase diagram and the factorizing curve can be detected.

Sec.\,\ref{sec:setup} contains a quick introduction to the model and the Bogoliubov circuits that connect different ground states. In Sec.\,\ref{sec:momentum-space} we derive the momentum space complexity, followed by the real space complexity in Sec.\,\ref{sec:real-space}. Sec.\,\ref{sec:penalty} shows how penalty factors enables us to detect more features of the phase diagram. Sec.\,\ref{sec:scaling} concerns the scaling limit close to the phase transition.

\section{Setup} \label{sec:setup}
The Hamiltonian for the XY chain is given by
\begin{equation}
    H=-\frac{1}{2}\sum_{j=1}^{N}\left[\left(\frac{1+\gamma}{2}\right)\sigma_{j}^{x}\sigma_{j+1}^{x}+\left(\frac{1-\gamma}{2}\right)\sigma_{j}^{y}\sigma_{j+1}^{y}+h\sigma_{j}^{z}\right].
\end{equation}
Following the procedure in \cite{franchiniIntroductionIntegrableTechniques2017a}, the Hamiltonian can be diagonalized by a Jordan-Wigner transformation giving
\begin{equation}
    H = h\sum_{j=1}^N \psi^\dag_j \psi_j -\frac{1}{2} \sum_{j=1}^{N-1} \left(\psi_j^\dag \psi_{j+1} +  \gamma \psi_j^\dag \psi_{j+1}^\dag+ h.c. \right) 
\end{equation}
 followed by a Bogoliubov rotation to give
 \begin{equation}\label{bh}
    H= \frac{1}{N}\sum_{q\in \Gamma}^{}\epsilon(q) \chi^{\dagger}_{q}\chi_{q} 
\end{equation}
where $\epsilon(q)=\sqrt{(h-\cos q)^2 + (\gamma \sin q)^2}$ and $\Gamma=\frac{2\pi}{N}\left(-\frac{N-1}{2},-\frac{N-1}{2}+1,...,\frac{N-1}{2}\right)$. We have removed irrelevant energy shifts and boundary terms and as we will always work in the large $N$ limit. 

The phase diagram of this model is shown in Fig.\,\ref{fig:phase_diagram}. We will study the complexity of connecting the ground states of Hamiltonians with different parameters. We define $(\gamma_R,h_R)$ and $(\gamma_T,h_T)$ be the parameters corresponding to the reference and target states.
A unitary transformation that accomplishes this is
\begin{equation}\label{eq:unit}
    U=\prod_{q>0,q\in\Gamma}U_q=\exp \left[\sum_{q>0,q\in\Gamma}^{} (\nu_q^T - \nu_q^R) A_q\right]
\end{equation}
where 
\begin{align}
    A_q&=\frac{1}{N}\left(\chi^{\dagger}_q\chi^{\dagger}_{-q} +\chi_{q}\chi_{-q}\right) \\
    \nu_q&=\frac{1}{2}\arctan \left(\frac{\gamma \sin q}{h-\cos q}\right).
\end{align}
The fermions in $A_q$ can be chosen to be the Bogoliubov fermions of any state, as the functional form is invariant under Bogoliubov rotations.

In the following sections, we will evaluate the circuit complexity for different circuits, both in momentum space and real space. We will choose the reference and target states as ground states of the XY chain.%

\section{Momentum Space Complexity} \label{sec:momentum-space}
Because the different sectors $q$ are decoupled, we may consider them separately. 
Following Nielsen's geometric approach, we write the unitary as the end point of a path ordered exponential
\begin{align}
    U_q(s)=\overleftarrow{\mathcal{P}}\exp\left(\int_{0}^{s}ds' Y(s)A_q\right),
\end{align}
such that $U_q(0)=\mathds{1}$ and $U_q(1)=U_q$. We define the circuit depth as
\begin{equation}
    \mathcal{D}[U_q]=\int_{0}^{1}ds \left|Y(s)\right|^2,
\end{equation}
and to find the complexity $\mathcal{C}_q$ we minimize this quantity w.r.t. $Y$. 
The complexity is simply
\begin{equation}
    \mathcal{C}_q=\left|\Delta\nu_q \right|^2.
\end{equation}
The complexity for the full unitary is then the sum over each momentum mode:
\begin{equation}
    \mathcal{C}=\sum_{q}^{}\mathcal{C}_q=\sum_{q}^{}\left|\nu_q^R-\nu_q^T\right|^2
\end{equation}
When the system size goes to infinity, this is amenable to analytic treatment, but we save this treatment for the real space complexity in the next section. By Parseval's theorem, the momentum space complexity is a special case of the real space one.

\section{Real Space complexity}\label{sec:real-space}
\subsection{Transforming the circuit}
In this section, we reinterpret the momentum space circuit as a real space circuit. We define the real space complexity including penalty factors and analytically solve for the complexity. 
We use the same unitary but through a Bogoliubov rotation and Fourier transform we can express it in terms of the original fermions as
\begin{equation}
    U=\prod_{n=1}^{N-1}U_n=\prod_{n=1}^{N-1}\exp\left[K_nG_n\right],
\end{equation}
where 
\begin{align}
    K_n &= \frac{2i}{N} \sum_{q>0,q\in\Gamma}^{}\Delta\nu_q \sin(qn),\\
    G_n & =\sum_{j=1}^{N}i\left(\psi^{\dagger}_{j+n}\psi^{\dagger}_{j}-\psi_{j}\psi_{j+n}\right).
\end{align} 
The unitary effectively decouples different values of $n$, and thus the complexity of the unitary will be the sum of the complexity of unitaries corresponding to each $n$ mode. 

\subsection{Complexity}
We consider a path ordered exponential of the form
\begin{align}
    U_n(s)=\overleftarrow{\mathcal{P}}\exp\left(\int_{0}^{s}ds' \bar{Y}_n(s)G_n\right).
\end{align}
The circuit depth is 
\begin{equation}
    \mathcal{D}[U_n]=\int_{0}^{1}e^{n l} n^\beta \left|\bar{Y}_n(s)\right|^2 ds,
\end{equation}
where we include penalty factors $l$ and $\beta$. When these penalty factors are greater than zero, they punish the use of non-local gates.
The complexity for this mode is $\mathcal{C}_n = e^{n l} n^\beta \left|K_n\right|^2$ and the full complexity is just the sum
\begin{equation}\label{eq:complexity-sum}
    \mathcal{C}=\sum_{n=1}^{N-1}e^{2 n l} n^\beta\left| K_n\right|^2.
\end{equation}

We evaluate \eqref{eq:complexity-sum} in the infinite system size limit where
\begin{equation}\label{sum_to_int}
    I_n = \frac{2i}{N}\sum_{q>0,q\in\Gamma}^{}\nu_q\sin(qn) \rightarrow \int_{-\pi}^{\pi} e^{i q n}\nu_q dq.
\end{equation}
One can evaluate the above integral using contour integration techniques \citep{franchiniIntroductionIntegrableTechniques2017a, liuCircuitComplexityTopological2020a}. The answer to the integral is dependent on which phase we are in, and the details are shown in App.\,\ref{app:analytic-derivation}. In the disordered (D) and ordered (O) phase we have
\begin{align}
    I_n^{(D)}=\frac{i\pi}{2n}\left(\lambda_-^n-\lambda_+^{-n}\right) \\
    I_n^{(O)}=\frac{i\pi}{2n}\left(2-\lambda_+^n -\lambda_-^n\right) 
\end{align}
respectively, where
\begin{equation}\label{eq:lambda-solution}
    \lambda_{\pm}=\frac{h \pm \sqrt{h^2 + \gamma^2 -1}}{1+\gamma}
\end{equation}
These parameters are connected to the correlation lengths $\xi_{\pm}$ as
\begin{equation}\label{eq: correlation length}
    \xi_{\pm}=\frac{1}{|\ln \lambda_{\pm}|}
\end{equation}
The final result is 
\begin{equation}\label{penalty_C}
    \mathcal{C}=\sum_{n=1}^{\infty}e^{2 n l} n^\beta\left|I_n^T -I_n^R\right|^2,
\end{equation}
where one uses $I^{(O)}$ or $I^{(D)}$ depending on where the target and reference state are located. This sum can be performed and takes the schematic form
\begin{equation}
    \mathcal{C} = \sum_{i,j} \pm \operatorname{PolyLog}{\left(2-\beta , \lambda_i \lambda_j e^{2l} \right)},
\end{equation}
where we sum over those $\lambda$ which are contained in the unit circle. Note that this formula can also be used outside the radius of convergence of the sum in \ref{eq:complexity-sum}, where it represents an analytic continuation. However, it \textit{should} \textit{not} be used there, since the sum is the correct representation of the effort in constructing the circuit. 

\begin{figure}[th!]
    \centering
    \includegraphics[width=0.3\textwidth]{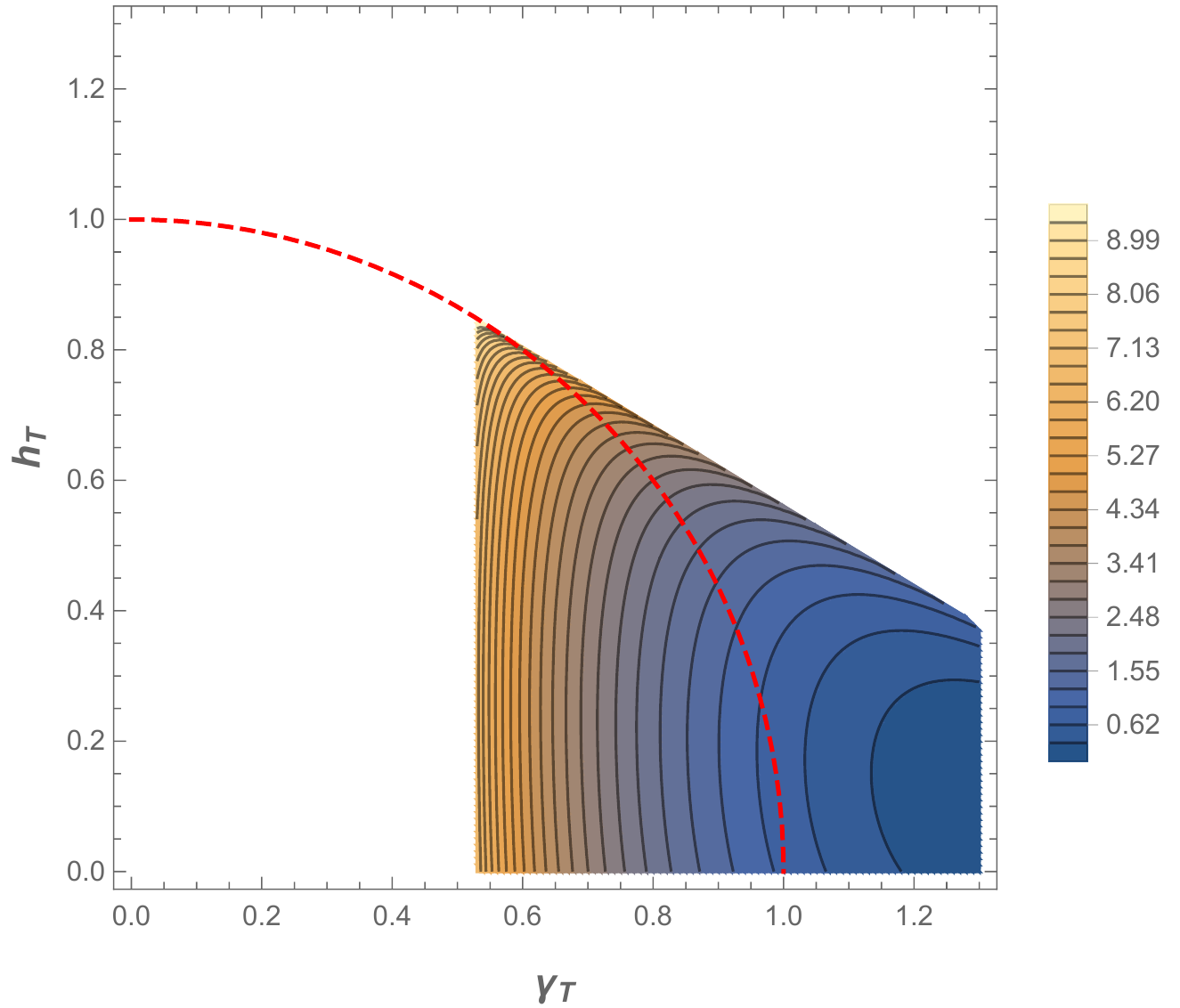}
    \caption{A contour plot of $\mathcal{C}$ as a function of $h_T$ and $\gamma_T$ with the parameters $l=|\ln 0.56|=0.58$, $h_R=0.1$, $\gamma_R=1.4$, $\beta=0$. The red dotted semicircle is factorizing curve. Complexity diverges in the white regions.}
    \label{fig:complexity_contours}
\end{figure}
\section{The effect of penalty factors}\label{sec:penalty}
In this section, we will study the influence of the penalty factors $l$ and $\beta$. For this section, we will assume that our reference state is in the ordered phase and far away from the phase transition. The case without penalty factors, $l=\beta=0$, is equivalent to the momentum space case by Parseval's theorem.

The penalty factor $e^{2 n l}$ penalizes the use of gates of length scales greater than $1/l$. When the correlation length of the target state exceeds this length scale, we should expect the complexity be large. In fact, the radius of convergence of the sum in \ref{eq:complexity-sum} is exactly where $l\xi = 1$. Beyond this point, the complexity is infinite. We see this explicitly in Fig.\,\ref{fig:complexity_contours}, which shows the complexity for a non zero value of $l$. The dashed curve is the factorizing curve and we note that the boundary features a sharp kink along this curve.

\begin{figure}[th!]
    \centering
    \includegraphics[width=0.3\textwidth]{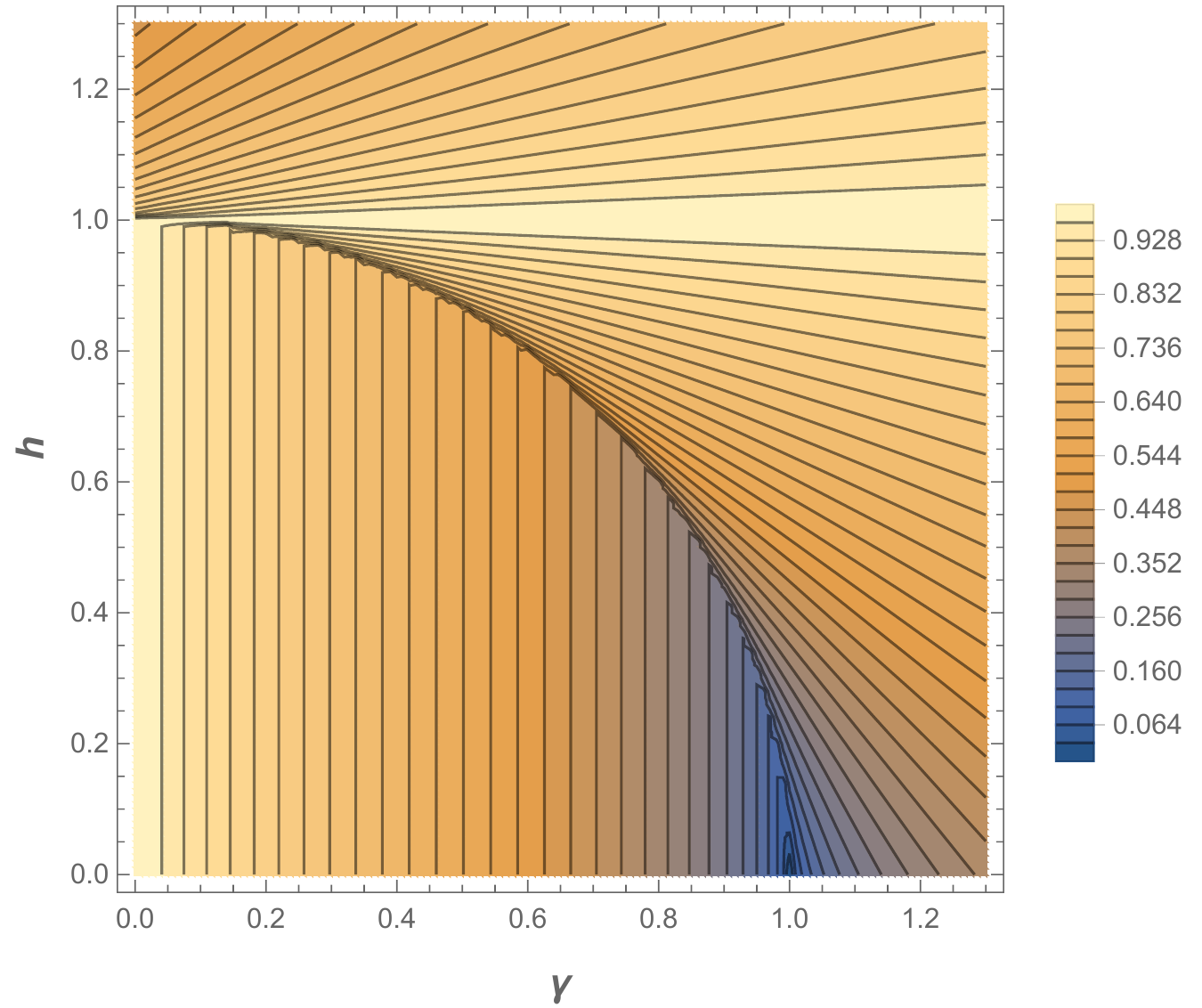}
    \caption{A contour plot of the relevant correlation parameters: $|\lambda_+|$ in the ordered phase and $|\lambda_+^{-1}|$ in the disordered phase as a function of $h$ and $\gamma$. Note that the 'oscillatory' phase is visible in this plot. }
    \label{fig:lambda_contours}
\end{figure}
For a given choice of $l$, the curve where $l\xi = 1$ divides states which are accessible from those which are not. We show these curves in a contour plot of $\lambda$ in Fig.\,\ref{fig:lambda_contours}. From this perspective, the appearance of the factorizing curve is clear. These are the points where $\lambda$ collides with its partner and develops an imaginary part. By introducing a penalty factor with a length scale, one can probe this feature.

\begin{figure}[t!]
    \centering
    \includegraphics[width=0.4\textwidth]{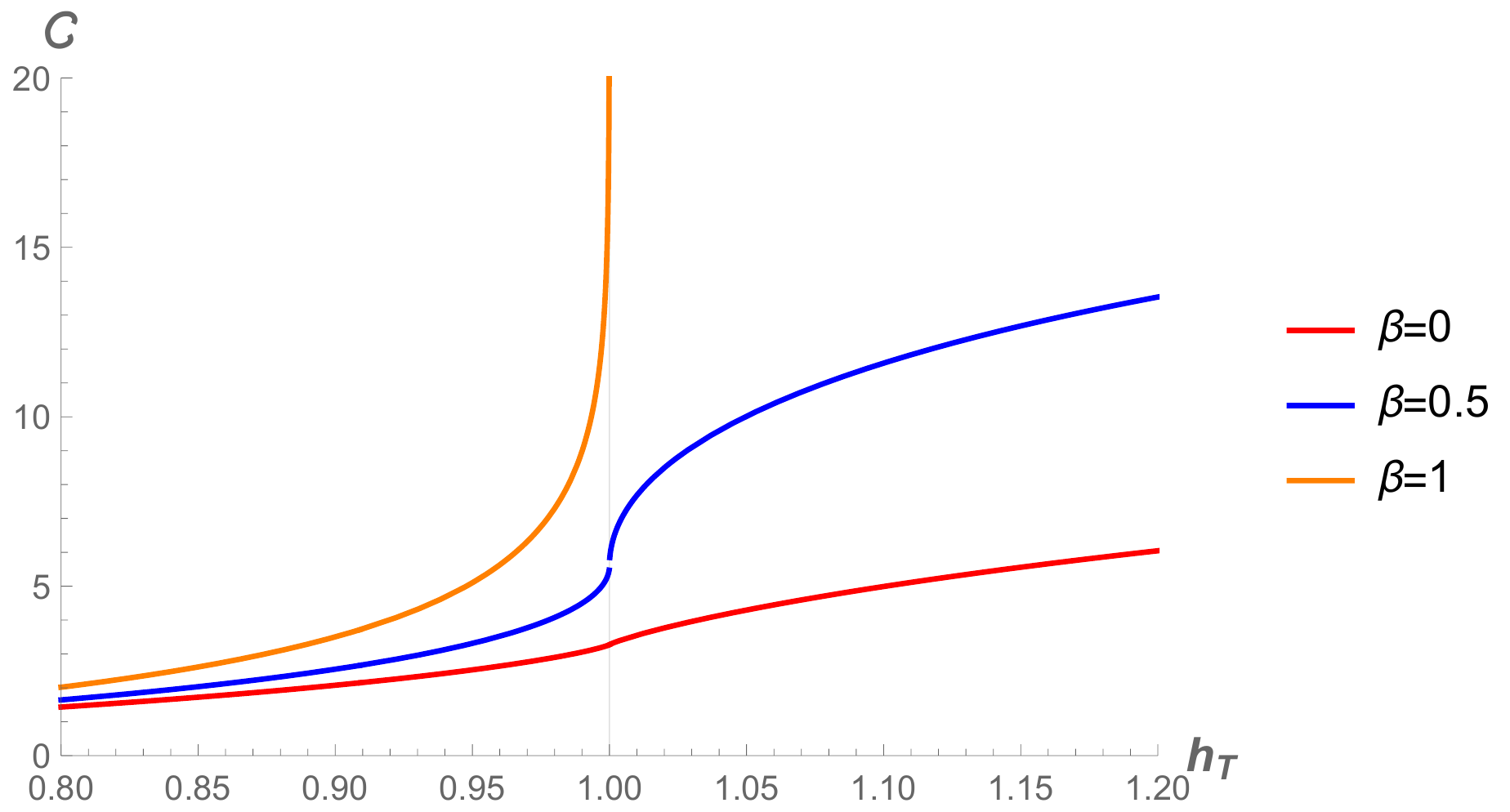}
    \caption{Circuit complexity as a function of $h_{T}$ for different values of $\beta$ with the parameters $h_{R}=0.1$, $\gamma_{R}=1.1$, $\gamma_{T}=1.1$, $l=0$. $\beta$ influences the analytic properties at the phase transition.}
    \label{fig:complexity_lineplot}
\end{figure}

We now come to $n^{\beta}$. This penalty factor penalizes the use of circuits with constituent gates of long length. A natural choice that models the case where circuit are built from local gates is $\beta=1$. To see this, note that the gates we use $G_n$ are built from fermions acting $n$ sites apart. One needs at least $n$ local gates to construct it, corresponding to $\beta = 1$.

The consequence of this penalty factor close to the phase transition is shown in Fig.\,\ref{fig:complexity_lineplot}. As $\beta$ is increased, the behaviour at the phase transition becomes sharper. At $\beta=1$, the complexity diverges at the transition. This matches the intuition that it takes an infinite number of local gates to reach the phase transition where the correlation length diverges. This is a complementary perspective to the momentum space picture where non-analyticities are interpreted as a consequence of the change in winding number \cite{liuCircuitComplexityTopological2020a}. 

\section{Scalings at the Phase transition}\label{sec:scaling}
In this section, we show analytic results for how complexity scales as the  target state approach the phase transition. We choose the reference state in the ordered phase far from the phase transition, $h_{T}=1-\epsilon$ and $\gamma_T\neq 0$.
We keep $l=0$, as otherwise, the phase transition is unreachable. The form of the scaling depends on whether $\beta$ is an integer less than or equal to 1 or not. The leading $\epsilon$ dependant terms in the relevant cases are
\begin{align}
    \mathcal{C}_{\beta=0}&\sim- \frac{\pi^2}{2\gamma_T}  \epsilon \ln \epsilon \\ 
    \mathcal{C}_{\beta=1}&\sim -\frac{\pi^2} {4}\ln \epsilon \\
    \mathcal{C}_{\beta \not\in \mathbb{Z}}&
    \sim  \frac{\pi^2}{2^{1+\beta}\gamma_T^{1-\beta}}\Gamma(\beta-1) \epsilon^{1-\beta}. 
\end{align}
When $\beta=0$, we see that $\mathcal{C}$ is finite, but its derivative $\pdv{\mathcal{C}}{h_T}$ diverges as $\ln \epsilon$, as has been noted before \citep{liuCircuitComplexityTopological2020a, xiongNonanalyticityCircuitComplexity2020,jaiswalComplexityInformationGeometry2021b, jaiswalComplexityInformationGeometry2022}. 
For $\beta\geq 1$, the complexity itself is a divergent quantity, as seen in Fig\,\ref{fig:complexity_lineplot}. We can interpret this in terms of the correlation length, which diverges as $\epsilon^{-1}$. If the cost of a gate is proportional to its length, and the correlation length of the system diverges, so does the complexity.
Apart from the case $\beta=1$, the coefficient must contain an energy scale to balance out the powers of $\epsilon$. This energy scale is set by $\gamma_T$, which determines the anisotropy of the spin chain. We also note that this leading term is independent of the reference state, whose influence only shows up at subleading orders in the expansion.

\section{Conclusion}\label{sec:conclusion}
We have reinterpreted the Bogoliubov circuits as real space circuits. This enabled us to study the effects of penalty factors that punish the use of non-local gates. Because of the simplicity of the circuit, we are able to find analytical results for the complexity even including penalty factors.

Our main result is that the complexity is sensitive to the factorizing curve in the XY model if penalty factors are included. This is evident in Figs.\,\ref{fig:complexity_contours} and \ref{fig:lambda_contours}. The physical reason is also clear from the analytical solution. By introducing a penalty factor with a length scale, we are able to probe the behavior of the correlation lengths of the system. These correlation lengths have a sharp change in behavior at this curve where they collide and develop an imaginary part, see Eq.\,\ref{eq:lambda-solution}.

The other penalty factor, $\beta$, influences how the complexity behaves at the phase transition. At $\beta = 0$, only the derivative of the complexity diverges, as noted in previous papers. With the choice $\beta=1$, which we argue models the case where the circuit is built from local gates, the complexity diverges. We show this numerically as well as analytically.

These findings show that including effects of locality, complexity is a more sensitive probe of the system than the momentum space formulation based on Bogoliubov circuits. We expect that the same should hold for a fully local formulation of complexity where the circuit is built from local gates.

\begin{acknowledgments}
We thank Michal P. Heller and Miłosz Panfil for their discussions and their involvement during the beginning of this project. NCJ is supported by funding from the Max Planck Partner Group Grant through Prof. Diptarka Das at IIT Kanpur. VS is supported by funding from the European Research Council (ERC) under the European Union’s Horizon 2020 research and innovation programme under Grant Agreement No. 856526.
\end{acknowledgments}

\nocite{*}
\bibliography{newcomplexitybib}
\appendix
\section{Fourier transforming the Bogoliubov circuit}\label{app:fourier}
Here we show how to reinterpret the Bogoliubov circuit
\begin{equation}
    U=\prod_{q>0,q\in\Gamma}U_q=\exp \left[\sum_{q>0,q\in\Gamma}^{} (\nu_q^T - \nu_q^R) A_q\right]
\end{equation}
as a real space circuit. As a first step, write the circuit in terms of the original momentum space fermions
\begin{equation}\label{eq:U_original_fermions}
   U=\exp\left[\sum_{q>0,q\in\Gamma}^{} \frac{\Delta\nu_q}{N}\left(\psi^{\dagger}_q\psi^{\dagger}_{-q} +\psi_{q}\psi_{-q}\right)\right].
\end{equation}
Now we rewrite the operator in the exponent in terms of the real space fermions
\begin{equation}
    \psi^{\dagger}_q\psi^{\dagger}_{-q} +\psi_{q}\psi_{-q}=i\sum_{j,k=1}^{N}\left(e^{iq(j-k)}\psi^{\dagger}_j\psi^{\dagger}_{k} + e^{-iq(j-k)}\psi_k\psi_{j}\right).
\end{equation}
We can rearrange the sum by defining $(j-k)=n$ as a new variable
\begin{equation}\label{eq:Aq_real_space}
    \psi^{\dagger}_q\psi^{\dagger}_{-q} +\psi_{q}\psi_{-q}=i\sum_{j=1}^{N}\sum_{n=1-N}^{(N-1)}\left(e^{iqn}\psi^{\dagger}_{j+n}\psi^{\dagger}_{j} + e^{-iqn}\psi_{j}\psi_{j+n}\right)
\end{equation}
Rewriting $e^{iqn}=\cos\left(qn\right) + i \sin \left(qn\right)$, we can rewrite the the above sum as
\begin{equation}
    \psi^{\dagger}_q\psi^{\dagger}_{-q} +\psi_{q}\psi_{-q}=i\sum_{n=-(N-1)}^{(N-1)}\left[\cos \left(qn\right)H_n + \sin \left(qn\right)G_n \right]
\end{equation}
where we have defined two new hermitian gates $H_n$ and $G_n$ as
\begin{align}
    H_n&=\sum_{j=1}^{N}\left(\psi^{\dagger}_{j+n}\psi^{\dagger}_{j}+\psi_{j}\psi_{j+n}\right)\\
    G_n&=\sum_{j=1}^{N}i\left(\psi^{\dagger}_{j+n}\psi^{\dagger}_{j}-\psi_{j}\psi_{j+n}\right)\\
\end{align}
By noticing that $H_n$ and $G_n$ are odd in $n$, it can be inferred that
\begin{equation}
    \sum_{n=-(N-1)}^{(N-1)}\cos \left(qn\right)H_n=0
\end{equation}
Plugging Eq.\,\ref{eq:Aq_real_space} into Eq.\,\ref{eq:U_original_fermions}, we find that the unitary connecting ground states can now be written as
\begin{equation}
   U=\exp\left[\sum_{n=1}^{N-1} K_n G_n\right],
\end{equation}
where $K_n = \frac{2i}{N}\sum_{q>0,q\in\Gamma}^{}\Delta\nu_q \sin(qn)$.
It can be shown that
\begin{equation}
    \left[G_n, G_{n'}\right]=0,
\end{equation}
and so the unitary can be factorized as
\begin{equation}
    U=\prod_{n=1}^{N-1}U_n=\prod_{n=1}^{N-1}\exp\left[K_nG_n\right].
\end{equation}

\section{Analytic derivation of complexity} \label{app:analytic-derivation}
We first construct the circuit for a given $n$ mode, and define the complexity for the same. We consider a path ordered exponential of the form
\begin{align}\label{}
    U_n(s)=\overleftarrow{\mathcal{P}}\exp\left(\int_{0}^{s}ds' \bar{Y}_n(s)G_n\right)
\end{align}
We minimize the circuit depth for this given unitary to obtain the optimal circuit. It is given by
\begin{equation}
    \mathcal{D}[U_n]=\int_{0}^{1}ds \left|\bar{Y}_n(s)\right|^2
\end{equation}
If we consider a circuit unitary ansatz of the form
\begin{equation}
    U_n(s)=\exp\left[a_n(s)G_n\right]
\end{equation}
The above unitary has the boundary conditions $a_n(0)=0$ and $a_n(1)=K_n$. For these set of boundary conditions the complexity is given by
\begin{equation}
    \mathcal{C}_n=\left|\frac{2i}{N}\sum_{q>0,q\in\Gamma}^{}\Delta\nu_q\sin(qn)\right|^2
\end{equation}
The complexity for the complete unitary will just be the sum of the complexities for each $n$ mode
\begin{equation}
    \mathcal{C}=\sum_{n=1}^{N-1}\mathcal{C}_n=\sum_{n=1}^{N-1}\left|K_n\right|^2
\end{equation}

\subsection{Fourier integral of $\nu_q$}
\begin{figure*}[t!]
    \centering
    \includegraphics[width=0.75\textwidth]{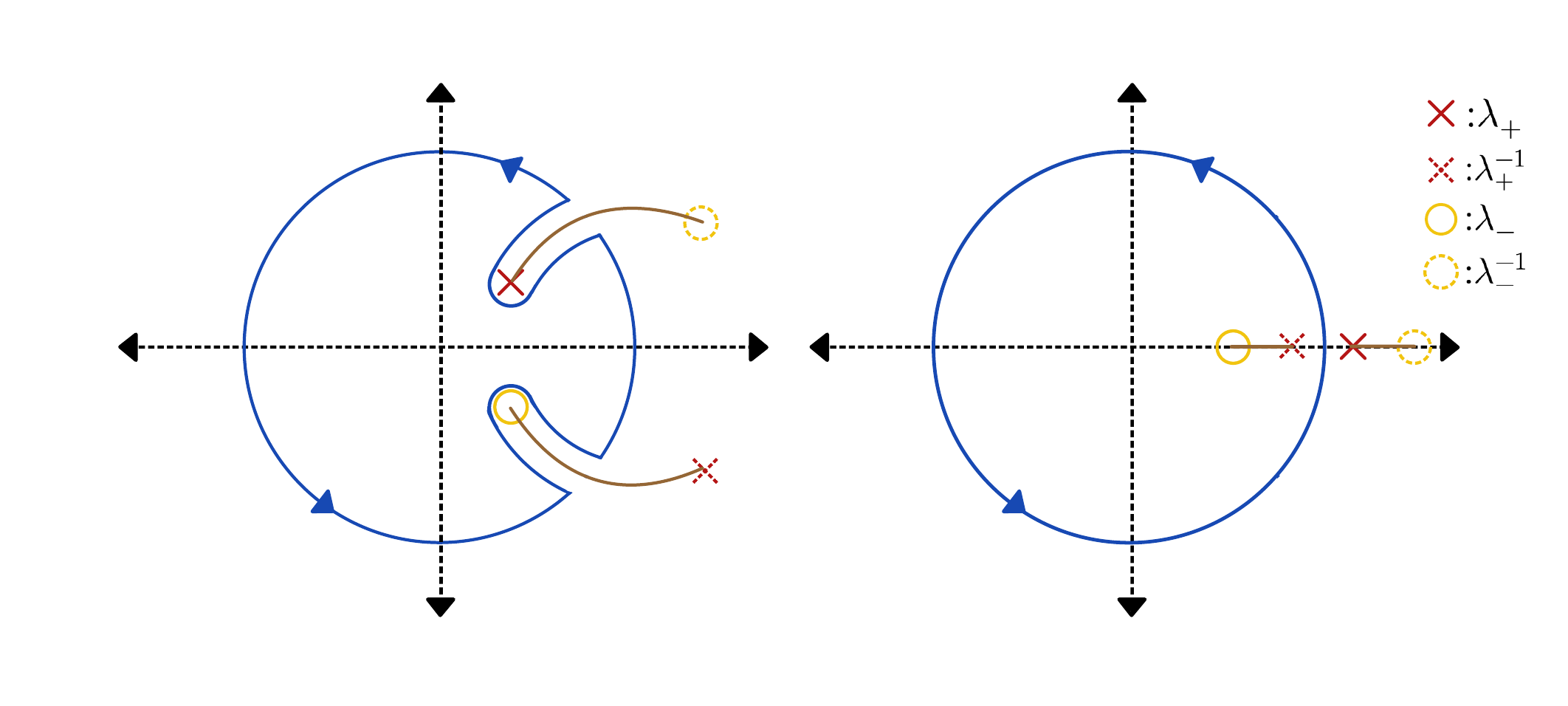}
    \caption{A complex plot showing the integration contour ($|t|=1$) in blue used to calculate the Fourier transform of $\nu_q$. The brown lines denote branch cuts. In the figure to the left the branch cut touches the integration contour at the points z and $\bar{z}$. The contour on the left is used for states which are in the ordered phase (within the oscillatory phase to be precise, but the branch cut leaves the contour in the same way for states outside the oscillatory phase but within the ordered phase, in which case $\lambda_+$ and $\lambda_-^{-1}$ move to the real line). The contour to the right is used for states in the disordered phase. As the contour completely surrounds the branch cut, one can deform it to be around the branch points within  to simplify the integral.}
    \label{integration contour}
\end{figure*}

To evaluate an explicit expression for the real space complexity, we are required to evaluate integrals of the form
\begin{equation}
    I_n=\int_{-\pi}^{\pi} e^{i q n}\nu_q  dq
\end{equation}
Making the change of variable $t=e^{i q}$, we have
\begin{equation}
    \frac{i}{2}\oint_{|t|=1} t^{n-1}\arctan\left[\frac{i \gamma  \left(t^2-1\right)}{-2 h t+t^2+1}\right]dt.
\end{equation}
Rewriting the arctan in terms of logarithms we have
\begin{equation}
    \frac{-1}{4}\oint_{|t|=1} t^{n-1}\log \left[\frac{\gamma +2 h t-(\gamma +1) t^2-1}{-\gamma +2 h t+(\gamma -1) t^2-1}\right]dt.
\end{equation}
One can evaluate the above integral using contour integration techniques, integrating over the unit circle contour\ref{integration contour}. The answer will be in form of the natural length parameters of the spin chain, given by
\begin{equation}
    \lambda_{\pm}=\frac{h \pm \sqrt{h^2 + \gamma^2 -1}}{1+\gamma}
\end{equation}

We can rewrite the integral in terms of these $\lambda$'s as
\begin{multline}\label{logexp}
    \frac{1}{4}\oint_{|t|=1} t^{n-1}f_{\lambda}(t) dt \\ = \frac{1}{4}\oint_{|t|=1} t^{n-1}\log \left[\frac{\left(t-\left(\frac{1}{\lambda }\right)_+\right) \left(t-\left(\frac{1}{\lambda }\right)_-\right)}{\left(t-\lambda _+\right) \left(t-\lambda _-\right)}\right]dt
\end{multline}
The answer to \eqref{logexp} will depend on which phase we are in. Studying the branch cut structure of $f_{\lambda}(t)$ will help us solve the integral.
\subsubsection{Disordered Phase}
\begin{figure}[H]
    \centering
    \includegraphics[width=0.35\textwidth]{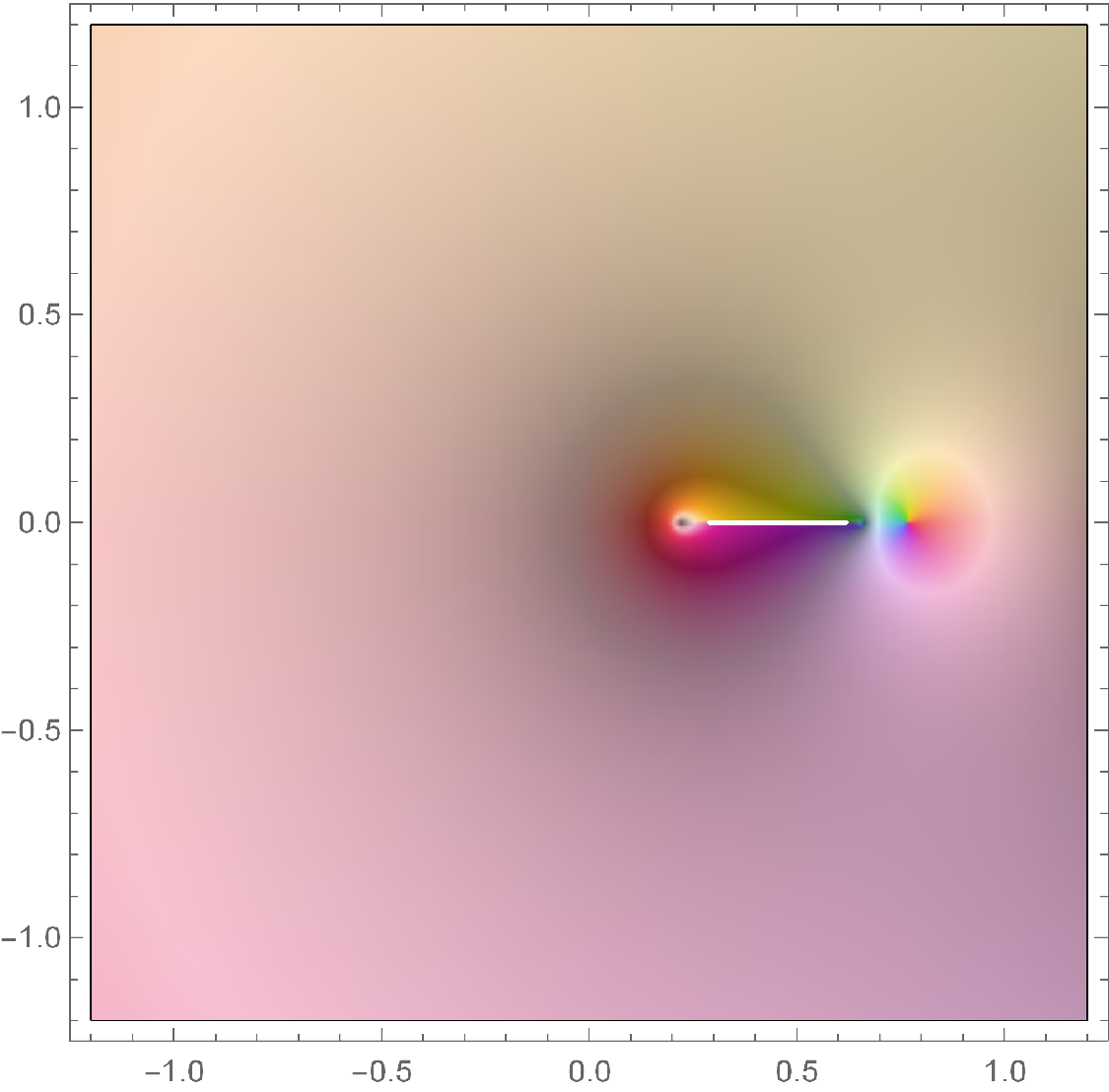}
    \caption{A complex plot of $f_{\lambda}(t)$ with $h\rightarrow 1.3$, $\gamma\rightarrow0.5$, which is in the disordered phase. The white line in the plot is a branch cut that connects $\lambda_-$ on the left to $\lambda_{+}^{-1}$ on the right. On crossing the branch cut from the magenta side to the green side, one picks up a constant piece of $2\pi i$ due to the logarithm.}
    \label{integration contour disordered}
\end{figure}

In the disordered phase, the branch cut is completely contained within the unit contour, we can deform the contour to be around the branch cut and solve the contour integral. The contour that goes around the branch point goes to zero as there is no pole, and we are only left with the integrals above and below the branch cut
\begin{multline}
    \frac{1}{4}\oint_{|t|=1} t^{n-1}f_{\lambda}(t)dt \\ =\frac{1}{4}\left[\int_{\lambda_{+}^{-1}}^{\lambda_-}t^{n-1}(2\pi i +f_{\lambda}(t))dt+\int_{\lambda_-}^{ \lambda_{+}^{-1}}t^{n-1}f_{\lambda}(t)dt\right] \\= \frac{i \pi}{2}\int_{\lambda_{+}^{-1}}^{\lambda_-}t^{n-1}dt
\end{multline}
Thus, in the disordered phase
\begin{equation}
    I_n=\frac{1}{4}\oint_{|t|=1} t^{n-1}f_{\lambda}(t)dt=\frac{i\pi}{2}\left(\frac{\lambda_-^n}{n}-\frac{\lambda_+^{-n}}{n}\right)
\end{equation}

\subsubsection{Ordered Phase}
\begin{figure}[H]
    \centering
    \includegraphics[width=0.35\textwidth]{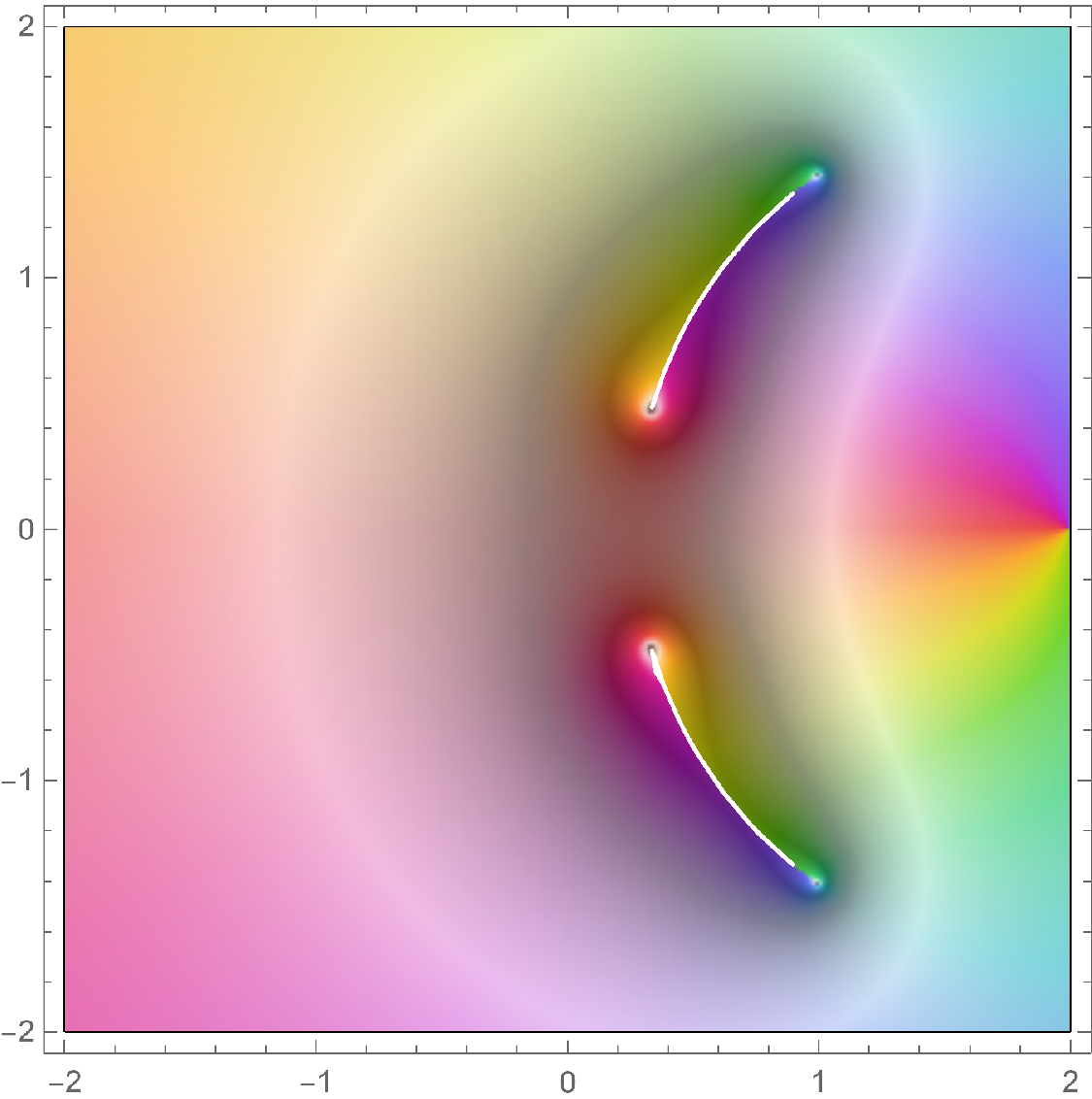}
    \caption{A complex plot of $f_{\lambda}(t)$ with $h\rightarrow 0.5$, $\gamma\rightarrow0.5$, which is in the ordered phase. The white lines in the plot are branch cut which connects  $\lambda_{+}$ on the left to $\lambda_{-}^{-1}$ on the right (top half), and connects $\lambda_-$ on the left to $\lambda_{+}^{-1}$ on the right (bottom half). On crossing the branch cuts from the magenta side to the green side, one picks up a constant piece of $2\pi i$ due to the logarithm.}
    \label{integration contour ordered}
\end{figure}
In the ordered phase, the branch cuts move out of the unit contour, and we will denote the point at which the branch cut crosses the unit circle in the top half plane as $z$. We can use the inherent mirror symmetry to come to the conclusion that $\bar{z}$ will be the point at which the branch cut crosses the unit circle in the lower half plane. In such a case, we would need to deform our contour like the left figure in \ref{integration contour}. Once again, the integrals around the branch points will turn out to be zero, and we are left with the integrals above and below the branch cuts
\begin{multline}
    \frac{1}{4}\oint_{|t|=1} t^{n-1}f_{\lambda}(t)dt  =\frac{1}{4}\Big[\int_{\lambda_{+}}^{z}t^{n-1}(2\pi i +f_{\lambda}(t))dt \\ +\int_{z}^{\lambda_{+}}t^{n-1}f_{\lambda}(t)dt + \int_{\lambda_{-}}^{\bar{z}}t^{n-1}(2\pi i +f_{\lambda}(t))dt \\+ \int_{\bar{z}}^{\lambda_{-}}t^{n-1}f_{\lambda}(t)dt\Big]
\end{multline}
This integral simplifies to
\begin{equation}
    \frac{1}{4}\oint_{|t|=1} t^{n-1}f_{\lambda}(t)dt=\frac{i \pi}{2}\int_{\lambda_{+}}^{z}t^{n-1}dt + \frac{i \pi}{2}\int_{\lambda_{-}}^{\bar{z}}t^{n-1}dt
\end{equation}
Thus, in the ordered phase
\begin{equation}
    I_n=\frac{i\pi}{2}\left(\frac{z^n}{n}+\frac{\bar{z}^n}{n}-\frac{\lambda_+^n}{n}-\frac{\lambda_-^n}{n}\right)
\end{equation}
Since $z$ is on the unit circle, we can take $z\rightarrow e^{i \theta}$ and rewrite $I_n$ as
\begin{equation}
    I_n=\frac{i\pi}{2}\left(\frac{e^{i n \theta}}{n}+\frac{e^{-i n \theta}}{n}-\frac{\lambda_+^n}{n}-\frac{\lambda_-^n}{n}\right)
\end{equation}

While we have put the two branch cuts at $\theta$ and $-\theta$, in order to get a correct circuit, they need to coincide. At a branch cut, the Bogoliubov angle jumps by $\pi/2$. But the circuit is only invariant (up to a sign) under shifts by $\pi$, and so these two shifts have to happen at the same point. Therefore we put $\theta = 0$.

\end{document}